\documentclass[11pt,preprint]{aastex}
\newcommand{\be}{\begin{equation}}
\newcommand{\ba}{\begin{eqnarray}}
\newcommand{\ee}{\end{equation}}
\newcommand{\ea}{\end{eqnarray}}

\def\lsim{\mathrel{\mathstrut\smash{\ooalign{\raise2.5pt\hbox{$<$}\cr\lower2.5pt\hbox{$\sim$}}}}}
\def\gsim{\mathrel{\mathstrut\smash{\ooalign{\raise2.5pt\hbox{$>$}\cr\lower2.5pt\hbox{$\sim$}}}}}

\def \se        {\!=\!}

\begin{document}

\title{An X-ray Galaxy Cluster Survey for Investigations of Dark Energy \\
a White Paper submitted to the Dark Energy Task Force, 15 June 2005 \\
{\small point of contact: keith.m.jahoda@nasa.gov}}

\vspace{-\baselineskip}
{\small \it
Z. Haiman$^f$, 
S. Allen$^r$,
N. Bahcall$^q$,
M. Bautz$^l$, 
H. Boehringer$^n$,
S. Borgani$^s$,
G. Bryan$^f$,
B. Cabrera$^r$,
C. Canizares$^l$, 
O. Citterio$^b$, 
A. Evrard$^o$,
A. Finoguenov$^n$, 
R. Griffiths$^d$,
G. Hasinger,$^n$ 
P. Henry$^i$, 
K. Jahoda$^h$, 
G. Jernigan$^a$, 
S. Kahn$^r$,
D. Lamb$^e$,
S. Majumdar$^c$,
J. Mohr$^j$, 
S. Molendi$^t$,
R. Mushotzky$^h$, 
G. Pareschi$^b$,
J. Peterson$^r$,
R. Petre$^h$, 
P. Predehl$^n$,
A. Rasmussen$^r$,
G. Ricker$^l$,
P. Ricker$^k$, 
P. Rosati$^g$,
A. Sanderson$^j$,
A. Stanford$^k$,
M. Voit$^p$,
S. Wang$^f$,
N. White$^h$,
S. White$^m$
}

{\footnotesize
$^a$Berkeley,
$^b$INAF-OAB,
$^c$CITA,
$^d$Carnegie Mellon,
$^e$Chicago,
$^f$Columbia,
$^g$ESO,
$^h$GSFC,
$^i$Hawaii,
$^j$Illinois,
$^k$Livermore,
$^l$MIT,
$^m$MPA,
$^n$MPE,
$^o$Michigan,
$^p$Michigan State,
$^q$Princeton,
$^r$Stanford,
$^s$Trieste,
$^t$IASF-INAF
}

The amount and nature of dark energy (DE) can be tightly constrained
by measuring the spatial correlation features and evolution of a sample of $\sim 100,000$ galaxy clusters over the redshift range $0<z\lsim 1.5$.
Such an X-ray survey will discover {\it all} collapsed structures with mass above
$3.5\times 10^{14} h^{-1}{\rm M_\odot}$ at redshifts $z<2$ (i.e. the full
range where such objects are expected) in the high 
Galactic latitude sky.  Above this mass threshold the tight correlations
between X-ray observables and mass allow direct interpretation of the data.

DE affects both the abundance and the spatial
distribution of galaxy clusters.  Measurements
of the number density $d^2N/dMdz$ and the three--dimensional power
spectrum $P(k)$ of clusters are complementary 
(have different parameter degeneracies) to other DE probes, such as
Type Ia SNe or CMB anisotropies, and precisely constrain
cosmological parameters.

The abundance $dN/dz$ and power spectrum $P(k)$ of collapsed dark
matter halos are theoretically computable from ab--initio models,
with no free parameters (other than cosmology). 
Uncertainties in the relation between the
halo mass and the observable X-ray flux can be overcome through
a process of self-calibration, taking advantage of the synergy   
between the two observables.
While clusters are highly biased tracers of the mass distribution, 
the bias is calculable from the same simulations that derive the mass
function.  Hence the large bias is a
bonus - it increases the signal--to--noise of the $P(k)$ measurement
by a (mass limit dependent) factor of 10-100.

X-ray emission is an efficient and robust way to identify clusters.
Imaging X-ray cluster surveys have high and well understood completeness, low
rates of contamination, and
the selection function is well understood without complex simulations.  

The DE investigations that we describe can be performed
with a survey of 20,000 deg$^2$ to a 0.5--2 keV flux limit of 
$2.3 \times 10^{-14} ~{\rm erg~cm^{-2}~s^{-1}}$.  
At this flux the X-ray sky is dominated by clusters and AGN,
which can be separated with an angular resolution of 15 arsec.
The number--flux
relationship is well known to the proposed depth
(Gioia et al. 2001; Rosati et al. 2002).
The proposed survey, consistent in technical scope with
a NASA Medium Explorer mission, will identify $\sim 100,000$ clusters.
Multi-band optical surveys to provide the required photometric redshifts are already
in the planning stages, 
and will be contemporaneous with or precede our X-ray survey.

\clearpage

\section*{b. Precursor Observations}

We need accurate cluster redshifts ($\sigma_z \sim 0.02$) 
to $z \sim 1.5$ for $>$ 100,000 clusters. Most of these redshifts will 
be derived from multi-band photometry.
Photometric redshifts from recent 
optical surveys (Csabai et al. 2003; Blindert et al. 2004; 
Hsieh et al. 2005) have
demonstrated this level of
statistical precision to redshifts $z \sim 0.6-1$.
Deeper surveys
(itemized  below) will provide
photometry of similar accuracy to redshift $z \sim 1.5$.
With large spectroscopic training
sets
systematic biases can be controlled at the level of $\delta z\sim 0.001$ (see the 
Dark Energy Survey (DES) white paper submitted to this same panel), which is small enough so
that it makes no meaningful contribution to the error budget.

There are two key considerations for photometric cluster redshift determination. 
First, the redshift signal comes from shifting the 
prominent 4000 Angstrom break in red early-type cluster galaxies through the various bands, thus the 
optical photometry must extend to a wavelength of 10,000  Angstrom for clusters at z = 1.5. Second, the 
photometry must be deep enough to obtain redshifts for $\sim 10$ cluster members in order to reduce 
the statistical errors on any one galaxy.

Planned surveys meet both requirements.
The Sloan Digital Sky Survey (SDSS) exists and provides spectroscopic 
redshifts of cluster galaxies to $z \sim 0.6$ over $\sim 7000$ deg$^2$. 
The photometric redshifts derived from the 5 SDSS bands are accurate to 
0.03 for an individual galaxy. The 
PanSTARRS survey instrument is currently under construction in Hawaii. It will survey 30,000 
deg$^2$ in five bands (g, r, i, z, and Y, i.e. to 10,000  Angstrom)  to sufficient depth to provide 
redshifts of the required accuracy to z = 1.5. The DES, under consideration by 
the DOE and NSF, will survey 5000 deg$^2$ in the south, complementary to PanSTARRS, in 
four bands (g, r, i, and z) to sufficient depth to provide redshifts of the required accuracy to 
$z \ge 1.3$. 
The Large Synoptic Survey Telescope (LSST) is currently being designed. It will provide five band 
data much deeper than the required limit for a solid angle approaching 20,000 deg$^2$. 
Training sets will be available from public spectroscopic data, such as SDSS, 2dF-SDSS, 
VIRMOS-VLT Deep Survey and the Keck Deep2 Survey.

The DES survey complements (by design) the CMB mapping experiment by the
South Pole Telescope (SPT) which is funded to begin operation in 2007 
and which will deliver a sample of more than
20,000 galaxy clusters over 4000 deg$^2$ extending to
$z>1.5$.  Combining mm-wave
and X-ray data on clusters enables new science such as direct distance
measurements (Molnar et al. 2004).
The all-sky, low angular resolution (5-10
arcminute) CMB mapping experiment Planck  will deliver
interesting SZE observations of many massive and nearby galaxy clusters.

\section*{c. Systematic Errors }

The primary systematic concerns in a cluster survey fall into four
categories: (1) cluster mass uncertainties, (2) cluster sample
completeness and contamination, (3) theoretical uncertainities and (4)
redshift accuracy (discussed in ``Precursor 
Observations").  In fact, analyses of current
cluster samples of a few hundred systems are already systematics
limited in their constraints on $\Omega_M$ and $\sigma_8$ because of
uncertainties in galaxy cluster masses (e.g. Pierpaoli et al. 2003;
Evrard et al. 2002; Schuecker at al. 2003; Henry 2004).
Self-calibration (below) and external calibration (e.g. from detailed
studies of numerous individual clusters with Constellation--X, see the white
paper submitted to this panel) will address these systematics.
Below we
discuss each systematic in turn.

{\it Cluster masses:}  During the matter dominated phase that fosters the
growth of large-scale structure, clusters lack a clearly defined edge
(Lacey  \& Cole 1993; White 2001; Busha et al 2005).  Still, the halo
population can be
ordered by a mass $M_\Delta$ contained within a characteristic density
$\rho \se \Delta \rho_{crit}(z)$.  Correlations between $M_\Delta$ and
bulk observables like X-ray luminosity or temperature are exhibited in
both observations (Mohr et al. 1999;  Voevodkin \&
Vikhlinin 2004; Ettori et al 2004; Voit 2005) and hydrodynamical
structure formation
simulations (Bryan \& Norman 1998;  Mathiesen \& Evrard 2001; Gardini
\& Ricker 2004; Kravtsov et al. 2005;  Motl et al. 2005).  The
mass-observable correlations possess
a known intrinsic scatter that must be included in any cosmological analyses 
(Levine et al. 2002; Lima \& Hu 2005).

In a large cluster survey there are several reservoirs of
information about cosmology and cluster structure (including mass).  
These include the redshift
distribution of clusters, their spatial clustering, their intrinsic
shapes and alignments, the forms of the luminosity and temperature
functions as function of redshift, as well as scaling relations between various bulk cluster properties.  In a recent breakthough, it has
been shown that the information in deep surveys is rich enough to
solve for the unknown mass-observable parameters with only modest
degradation of contraints on the nature of dark energy (Majumdar \&
Mohr 2003, 2004; Hu 2003; Lima \& Hu 2004; Wang et al 2004).  The
bottom line is that self-calibration of a large, clean cluster sample over large,
contiguous regions of the sky can overcome the
cluster mass uncertainties.
We emphasize that our analysis below does {\it not} assume that the cluster 
mass can be computed from any ab--initio model.

The sensitivity of a self-calibrated cluster survey is greatly
improved 
if a small fraction of the clusters have externally 
derived masses
(Majumdar and Mohr 2004).  
Such external calibration could come 
from deep imaging and spectroscopic X-ray measurements of selected objects 
(Chandra and XMM will provide a few hundred objects; Constellation-X will
eventually provide a few thousand objects; many of which will be
identified by this survey).
Additionally,  the optical datasets which provide redshifts also
provide shear maps that, when appropriately averaged, offer an
independent calibration tool for the mass--luminosity and
mass--temperature  relations.   
While deep optical maps
of the survey region could in principle be used to define a shear selected cluster
sample, such samples are expected to have low ($<50 \%$)
completeness and high contamination, with at least as many false detections
as real clusters (Hamana et al. 2004; Hennawi \& Spergel 2005).  These
attributes currently limit the value of shear selected surveys for
cluster cosmology.  However, the statistics of the full shear map, or
cosmic shear, offers additional cosmological constraints
(Bacon et al. 2001; 
Jain 2002;
Bernstein \& Jain 2004; Massey et al. 2005).

{\it Cluster selection:} 
Cluster selection must be well understood to use the full statistical power of a 100,000 cluster sample.
Cluster survey completeness and contamination are well understood in the
X-ray regime.  Because of the high contrast of cluster X-ray emission
relative to the background, 
selecting clusters requires characterizing X-ray sources as extended
(clusters or galaxies) or point-like (AGN and stars).  
Clusters outnumber galaxies by a factor of several hundred;  X-ray colors
and optical counterparts separate these populations.
Mock observations
including clusters and AGN
indicate that $>90$\% of clusters at the proposed flux limit (corresponding
to 50 photons)
are recovered with a 15 arcsec HEW imager.
In addition, the contamination of these same samples is less
than a few \%, consistent with previous X-ray 
surveys (Vikhlinin et al 1998;  Rosati et al. 1998).
For instance, the ROSAT-based 160 deg$^2$ survey (Vikhlinin et al 1998) 
had 223 X-ray extended objects of which
203 were cluster candidates after comparison with optical images.
201 were spectroscopically confirmed as clusters (Mullis et al. 2003)
giving a 1\% false positive rate for a survey using the much larger PSF (40 arcsec)
of the ROSAT PSPC.

To use the statistical power of $10^5$ clusters distributed over a range of redshift, 
we need to limit the effects of uncertainty in our survey completeness and contamination 
to the $\sim$1\% level.  Starting from such a high completeness and low contamination 
this kind of sample control is straightforward.  
In comparison to other cluster survey methods like the SZE, optical and
weak lensing, the X-ray selection is by far the cleanest approach for
selecting massive, collapsed halos (e.g. galaxy clusters).

{\it Theoretical Uncertainty:}
A remaining concern is the level of accuracy
in theoretical predictions of the mass function of collapsed objects
(e.g. Sheth \& Tormen 1999; Jenkins et al 2001; Hu \& Kravtsov 2003),
currently at the
$10\%$ level.  Although the very largest simulations (Springel et al. 2005)
remain expensive to produce, billion particle models will soon be
routine. 

Agreement between independent codes 
with increasing accuracy gives confidence in the reliability of the simulations.
The $10^{10}$ particle Millenium
simulation (Springel et al. 2005) which used a Gadget-2 tree code, produced a mass
function that agrees to better than $10\%$ with earlier work (Jenkins et al. 2001)
employing the Hydra
P$^3$M code (Couchman et al. 1995; MacFarland et al
1998).  For the virial scaling relation of dark matter halos,
agreement among five different N-body
simulation codes at the few percent-level is observed (Evrard 2004).
We expect space density calibration at the few percent-level or better
to be available at the time of survey analysis.


\section*{d. Results from an X-ray Cluster Survey - Expected Error Budget}

Understanding the dynamical characteristics of dark energy is almost
certainly
essential to understanding its nature. The survey we discuss will
provide
remarkably powerful and precise constraints on dark energy dynamics.
We have performed detailed forecasts for the uncertainties that will
be achieveable on DE parameters, taking into account
self--calibration.  The cosmological sensitivity is extracted from
$dN/dz$, the cumulative counts of clusters above a given X--ray flux,
and their distribution in redshift (in $\Delta z=0.05$ wide bins),
combined with measurements of $P(k)$ in wider ($\Delta z=0.2$) bins.
Note that $dN/dz$ represents a unique, exponential sensitivity to DE
through a combination of the comoving volume element $d^2V/dz d\Omega$,
and through the growth of fluctuations $g(z)$.  The power spectrum
contains cosmological information from the intrinsic shape of the
transfer function and also from baryon features (Blake \& Glazebrook
2003; Seo \& Eisenstein 2003, Linder 2003; Hu \& Haiman 2003).
Baryonic features (``wiggles'') have been included in our analysis
using KINKFAST (Corasaniti et al. 2004), a modified version of CMBFAST
(Seljak \& Zaldarriaga 1996), tailored for time-varying $w$. The
wiggles will be detectable at $\sim 3.5\sigma$ significance in 5
separate redshift bins, varying in width between $\Delta z=0.2-0.5$,
each containing $\approx 20,000$ clusters. Their use as ``standard rods'' 
account for roughly half of the $P(k)$ constraints on the DE (Hu \& Haiman 2003).
The depth of the survey also allows a measurement of the redshift
evolution of the $P(k)$ normalization, which is an independent, direct
assessment of fluctuation growth.

\begin{figure}
\includegraphics[width=1.0\columnwidth,angle=0.0]{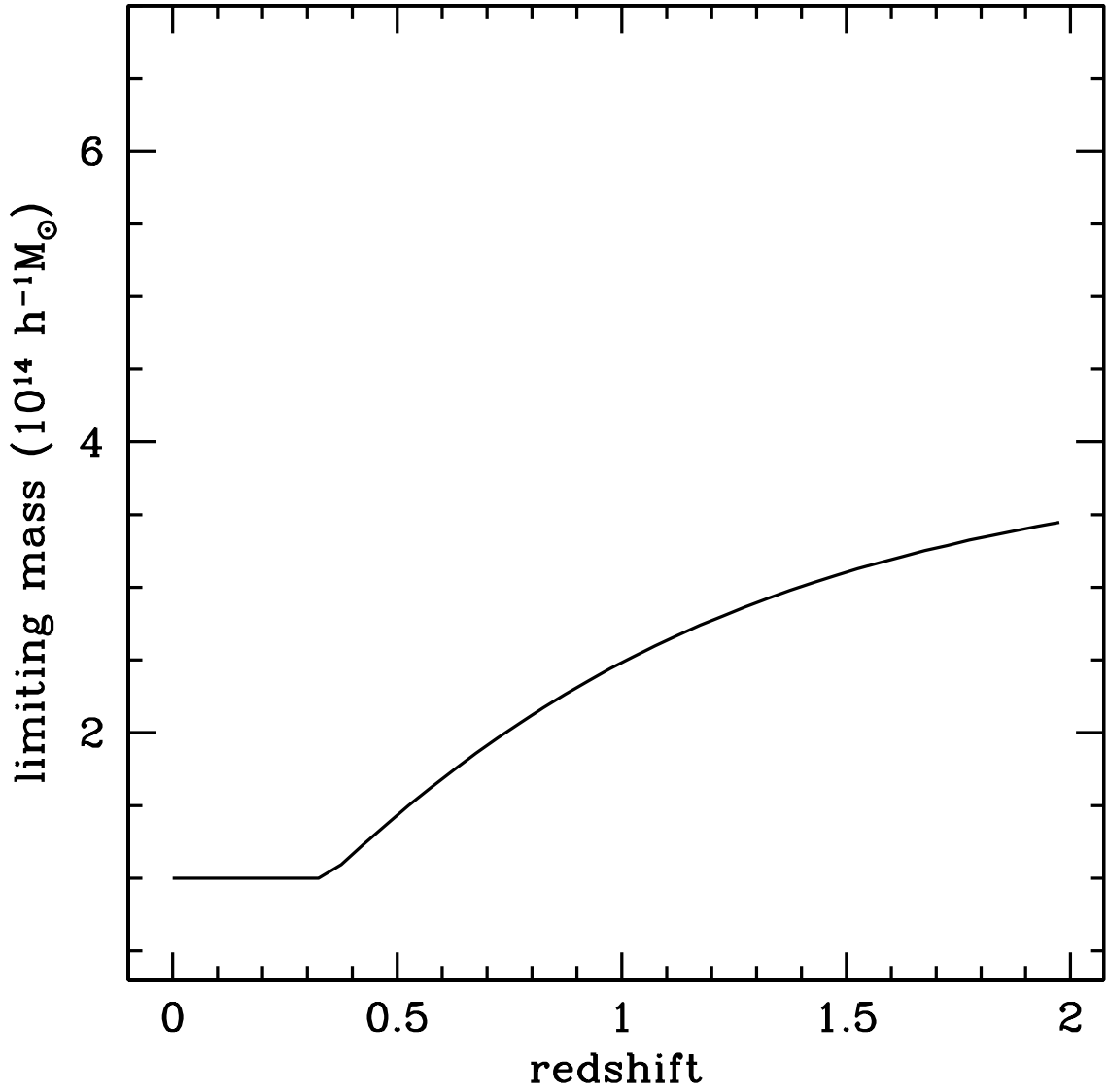}
\figcaption{The minimum detectable mass as a function of redshift for
a survey with a 0.5--2 keV flux limit of $2.3 \times 10^{-14}~{\rm erg}~{\rm cm}^{-2}~{\rm s}^{-1}$.
\label{mlim}}
\end{figure}

Figure \ref{mlim} shows the minimum mass of a detectable cluster in
the survey, corresponding to the $2.3 \times 10^{-14} {\rm
erg~cm^{-2}~s^{-1}}$ flux limit. At redshifts below $z<0.3$, the
smallest detectable objects have masses below $10^{14} h^{-1}{\rm
M_\odot}$ and we have ignored these objects (small groups, rather than
clusters) when deriving our constraints on DE.  The X--ray
survey will identify {\it all} clusters in the high latitude universe with masses
above $3.5\times 10^{14} h^{-1}{\rm M_\odot}$.

\begin{figure}
\includegraphics[width=1.0\columnwidth,angle=0.0]{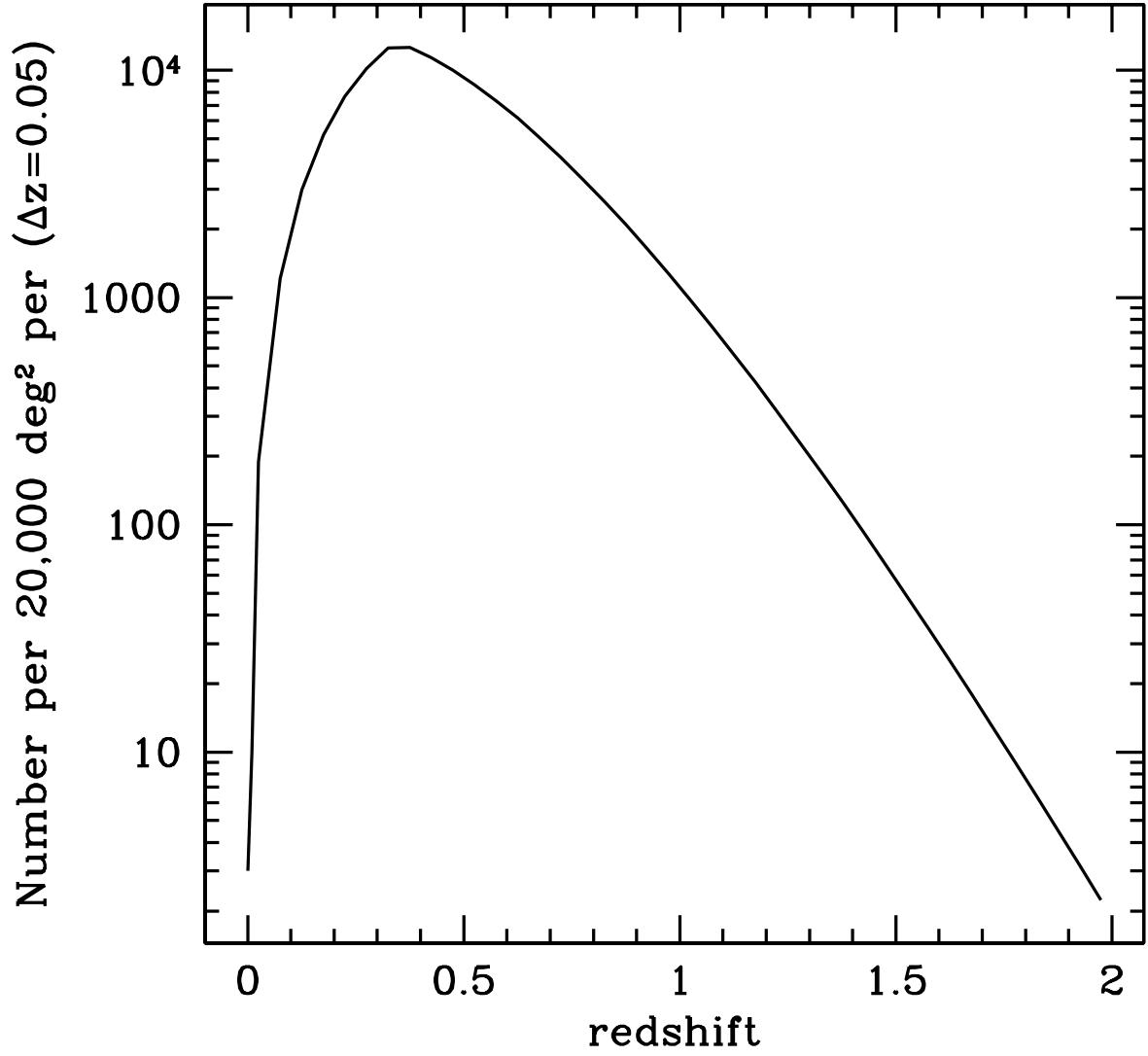}
\figcaption{Distribution in z of the clusters detected in the survey with 
$2.3 \times 10^{-14}~{\rm erg}~{\rm cm}^{-2}~{\rm s}^{-1}$ flux limit.
\label{zdist}}
\end{figure}

The expected redshift distribution (figure \ref{zdist}) has a mean
$\langle z \rangle=0.47$.
The survey is shallower than an SZE--selected sample (such as that
expected from SPT), and is comparable to a weak--lensing selected
sample (such as that expected from LSST). The redshift distribution
does have a significant tail out to high redshift, with $\approx 200$
clusters between $1.5<z<2$.  Given precise CMB measurements, the best
handle on DE properties is offered by the nearby (indeed,
$z=0$) clusters. High--$z$ clusters, however, significantly improve
constraints when CMB data are excluded.  This can also be
a powerful probe of DE in case its evolution is flatter than
expected so that DE already has significant dynamical effects at $1<z<2$ 
(to which the CMB data are insensitive).  High redshift clusters also allow a measurement of the evolution in the
normalization of the power spectrum within a single experiment
at the same scale, across the epoch of dark energy domination,
thus avoiding any uncertainty in the CMB-normalization and
its extrapolation to small scales.

We have utilized the Fisher information matrix to compute expected
1-$\sigma$ uncertainties on the DE density ($\Omega_{\rm DE}$), its
present--day equation of state ($w_0$) and its past evolution
($w_a$).  
We use $w_{a} \equiv dw/da$ (where
$a=(1+z)^{-1}$ is the scale factor).
Another common convention in the
literature is $w_{z} \equiv dw/dz$.  The uncertainty on $dw/dz$
would be a factor of $\sim$two smaller than we quote
below\footnote{This follows from Taylor-expanding $w(z)$ about $z =
0.5$, the redshift at which cluster survey sensitivity peaks.}.

We have incorporated 7 additional parameters in our analysis that the
cluster data are required to determine simultaneously with the 3 dark
energy parameters.  These include 4 additional cosmological parameters
for the baryon density ($\Omega_b h^2$), matter density ($\Omega_m
h^2$), and the power spectrum slope ($n_s$) and normalization
($\sigma_{8}$).  Self--calibration introduces 3 additional
non--cosmological parameters, which describe departures from the
expected dependence of the X--ray flux $f_x(M,z)$ on the cluster mass
$M$, and redshift $z$.  We assume a power--law relation of the form
$f_x(z) 4 \pi d_L^2 = A_x M^{\beta_x} E^2(z) (1+z)^{\gamma_x}$, where
$f_x$ is the flux limit, $d_L$ is the luminosity distance, $M$ is the
virial mass of the cluster, and $H(z) = H_0 E(z)$ is the Hubble
parameter at redshift $z$.

Table~\ref{bigsurvey} summarizes the 1-$\sigma$ uncertainties on the
dark--energy parameters, marginalized over the uncertainties of all
other parameters.  The top section summarizes self-calibrated models
while the bottom section gives the sensitivity assuming that the mass-observable 
relationship is reliably known externally. Constraints are evaluated around a spatially flat,
vanilla $\Lambda$CDM model. The first three rows use the full
10--parameter Fisher matrix, and show that strong constraints can be
obtained on the DE properties, including the evolution of the
equation of state (1$\sigma$ uncertainty on $w_a$ of 0.49),
despite the fact that we are requiring the survey to self--calibrate
the mass--flux relation.  The 2nd row in the table shows the
constraints available by combining the X--ray cluster survey with CMB
data.  The third row (with $w_a \equiv 0$) shows that the X-ray survey
is sensitive to $\Lambda$ models that depart from a pure
cosmological constant at the
$1\%$ level.
We assumed temperature and polarization anisotropy measurements
expected to be available from the Planck satellite in three frequency
bands (100, 143 and 217~GHz), with fractional sky coverage of $f_{\rm
sky} \approx 0.8$ (see Rocha et al. 2004 for details).  The addition
of Planck data pins down the geometry, distances, and power spectrum
normalization in the era before the DE dominated, and this further
reduces the uncertainty on $w_0$ and $w_a$ by approximately a
factor of 2 and 3, respectively. This underscores the complementarity
of cluster and CMB data in uncovering the nature of the DE.
The parameter degeneracies
arising from cluster constraints are highly complementary to those
from Type Ia SNe and the CMB (Wang \& Steinhardt 1998; Haiman et al. 2001; Holder
et al. 2001; Levine et al. 2002), as shown in figure
\ref{ellipses}.

As mentioned above, our Fisher analysis ignores several other
potential cluster observables that contain cosmological information,
such as the shape of the X--ray luminosity function, the number-flux
relationship, or scaling relations
between X--ray and other (SZE, optical) observables. Including this
extra information can, in principle, sharply reduce the degradation
of the constraints from self-calibration.  For reference, the last two
rows in Table~\ref{bigsurvey} therefore shows the constraints in a
perfect, idealized survey that does not require self--calibration.
This corresponds to using a 7--parameter cosmology--only Fisher matrix
that assumes that the $f_x$--mass relation has been calibrated to
$1\%$ accuracy (i.e., it assumes the self-calibration parameters are
known to $1\%$ precision).  The ultimate sensitivity of the X--ray
selected cluster sample to the equation of state parameter and its
evolution is an impressive 1$\sigma$ uncertainty of 0.01 and 0.066,
respectively.

In addition to studying the energy density and equation of state of
the DE, the X--ray cluster survey will be the ideal tool to
study any clustering of the DE.  In any model for the DE
other than the cosmological constant, the DE will cluster on large
enough scales (approaching the horizon scale). The best hope to detect
such clustering is through measuring the integrated Sachs--Wolfe
effect in the cross--correlation between CMB anisotropies and the two dimsnsional
angular power spectrum of a lower--redshift tracer of the
gravitational potential (such as galaxies; Hu \& Scranton 2004; Bean
\& Dore 2004). Note that the effects of the clustering on the CMB
power spectrum alone are hidden in cosmic variance. Clusters are as
good a tracer as galaxies for this test --- the main issues are sky
coverage and depth.  The number of tracers and their bias is less
relevant as long as the cluster power spectrum is also cosmic variance
limited.  Scaling from the results of Hu \& Scranton, we conclude that
in the example of a constant $w$=-0.8 model, the clustering of DE (in
quintessence, with sound speed $c_{\rm DE}$=1) is detectable at the
5-$\sigma$ level.

\section*{Context}

{\it e. Risks and Strengths:} A large contiguous sample of clusters traces the formation and growth
of structure in the universe and is a sensitive probe of DE.  An X-ray selected
sample of clusters is the most robust way to harvest a cluster sample with high completeness
and minimal contamination.   An X-ray sample can therefore be expected to reach higher statistical
precision before being limited by systematic concerns than other samples.  The risk,
demonstrated by current small surveys, is that the mass-observable relationship may
not be known to sufficient accuracy.  However, a proper survey overcomes this
limitation by measuring this relationship (self calibration).  Our proposal has
the very modest risk that the redshifts must be obtained externally.  However, the
{\it same} data required to obtain redshifts will be available for an independent calibration
of the mass via weak lensing which is a sensitive probe of total (baryonic plus dark) matter.

{\it f. Technology Readiness:}  Performing the cluster survey requires no major technology development. 
A detailed X-ray mirror design (Conconi \& Campana 2001) and 
a prototype mirror shell (Citterio et al. 1999) satisfy the
cluster survey mission requirements in table \ref{survey_description}.  
The baseline focal plane is an array of CCD detectors.
The underlying technology used for the XMM PN detectors (Meidinger et al. 2002),
Chandra ACIS CCDs (Burke et al. 1997; Bautz et al. 2004),
or the HETE Soft X-ray Camera (Villasenour et al. 2003) is both flight proven and acceptable.
A recent 6 month phase A study for a similar albeit smaller mission (DUO; Griffiths et al. 2004) demonstrated that there are no complicated requirements on the spacecraft or operations.  The survey concept is further elaborated in the appendix.

\begin{figure}
\includegraphics[scale=0.5,angle=-90.0]{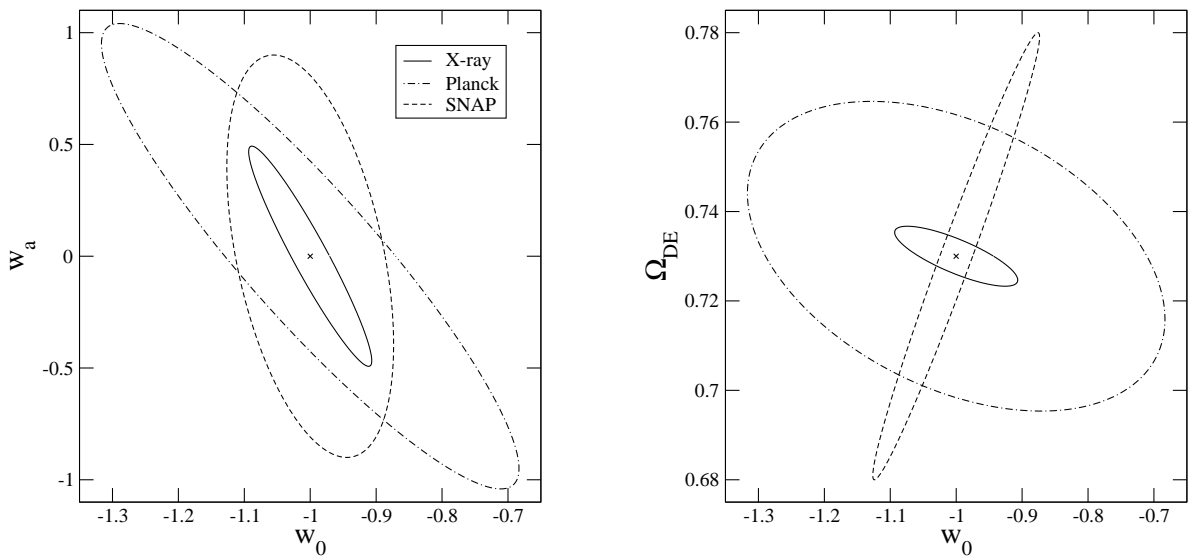}
\figcaption{Parameter constraints achieved with the self calibrated
X-ray survey, the Super Nova Acceleration Probe (SNAP), and Planck.  The
parameter degeneracies in the different experiments are highly
complementary demonstrating the synergy of the different investigations.  
The SNAP ellipse follows Aldering et al. (2004),
assumes a statistical uncertainty of 0.15 magnitude for each supernova,
a systematic error of $ 0.02*(1+z)/2.7$, and a number distribution
from their figure 9 (``final sample likely to be used") which results
in a total of $\sim 2000$ supernovae.
\label{ellipses}}
\end{figure}

{\it g. Relationship to JDEM and LSST:}
The X-ray survey is not a precursor to JDEM or LSST.  The substantially different parameter
degeneracies that result from a cluster survey and the light curves from
SNe make these techniques complementary (fig \ref{ellipses}).

{\it h. Access to Facilities:}
The X-ray survey needs access to photometric redshifts which can be determined from
observations which are already being planned.
Constellation-X
observations of high redshift clusters discovered with this mission
would provide complementary probes of cosmology and DE (see the Constellation-X white paper submitted
to this panel).

{\it i. Timeline:}
The large contiguous X-ray cluster survey considered here is not currently proposed 
to NASA.
The baseline mission is consistent with the resources (time, mass, volume, and cost) associated
with a NASA Medium Explorer mission, and as such could be flown in $\sim$ 2011.
Analysis techniques are known and straight-forward.

\clearpage

\clearpage

\begin{table}[tb]
\caption{Parameter uncertainties from a 100,000--cluster sample
(20,000 deg$^2$, $f_x= 2.3\times10^{-14}~{\rm erg~cm^{-2}~s^{-1}}$).
Results assume a spatially flat prior ($\Omega_{\rm DE}+\Omega_m=1$),
and have been marginalized over $\Omega_b h^2$, $\Omega_m h^2$,
$\sigma_8$, and $n_s$ (based on Wang et al. 2004).
\label{bigsurvey}}
\vspace{\baselineskip}
\begin{tabular}{|l|l|l|l|}
\hline
Self--Calibrated Experiment(s)             & $\sigma(w_0)$  & $\sigma(w_a)$ & 
$\sigma(\Omega_{\rm DE})$ \\
\hline
X-ray                              &  0.093       &   0.490  &   0.0067  \\
X-ray + Planck                     &  0.054       &   0.170  &   0.0052  \\
X-ray + Planck\tablenotemark{a}    &  0.016      &    -     &   0.0045  \\
\hline
Ideal Experiment\tablenotemark{b}             &            &    &  \\
\hline
X-ray                                         &  0.021       &   0.120  &   0.0030  \\
X-ray + Planck                               &  0.013       &   0.066  &   0.0027  \\
X-ray + Planck\tablenotemark{a}    &  0.0087      &    -     &   0.0019  \\
\hline
\end{tabular}
\tablenotetext{a}{assumes constant $w$ ($w_a = 0$)}
\tablenotetext{b}{Corresponds to  using a 7--parameter cosmology--only
  Fisher matrix that assumes that the $f_x$--mass relation has been
  calibrated to $1\%$ accuracy (i.e., it effectively assumes the
  self-calibration parameters are known to $1\%$ precision). }
\end{table}

\clearpage

\appendix
\section*{Appendix - Technical Implementation}

\begin{figure}
\includegraphics[scale=0.85,angle=0.0]{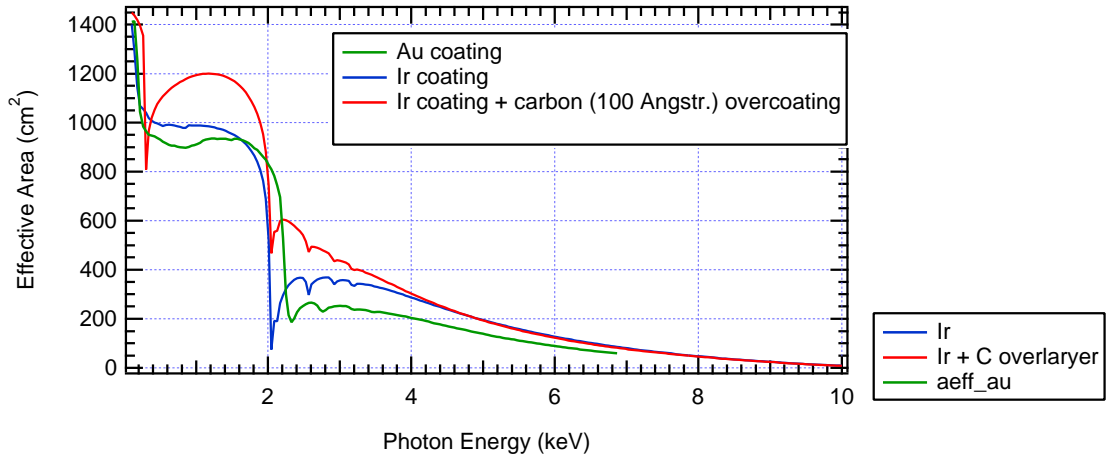}
\figcaption{The expected effective area of one WFXT telescope module
for various surface coatings.
\label{aeff}}
\end{figure}

\begin{figure}
\includegraphics[scale=0.55,angle=0.0]{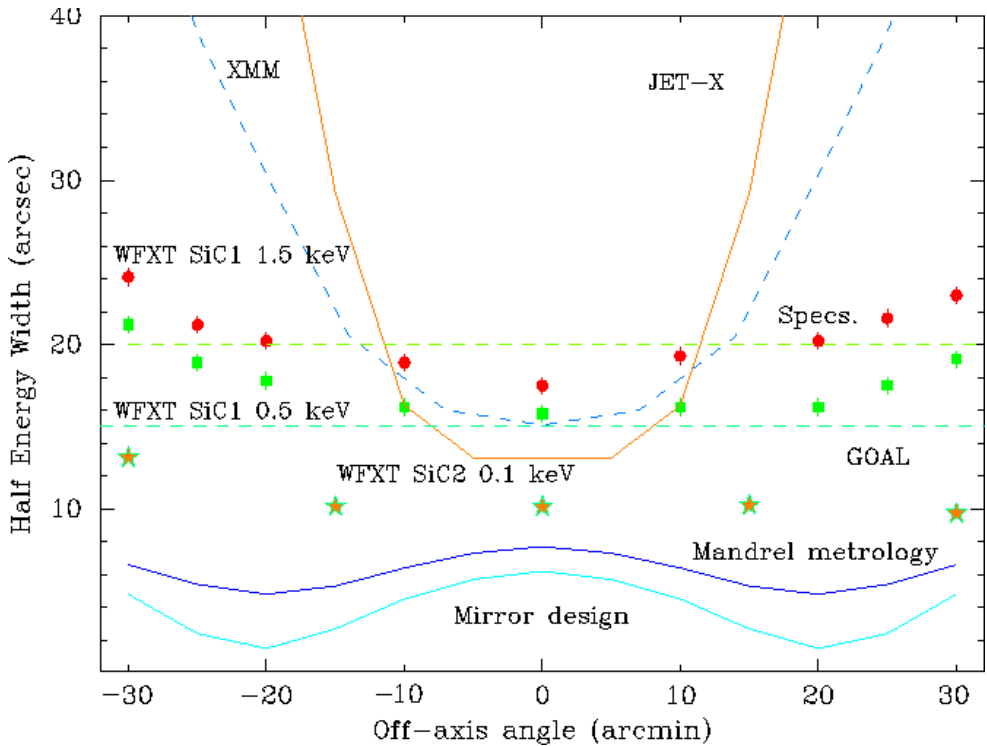}
\figcaption{The symbols show the measured performance of a single mirror shell (Citterio et al. 1999).  The performance at higher energies is degraded
due to roughness of the forming mandrel.  Significant improvement
is already demonstrated (on-axis) for a mirror shell fabricated
after re-polishing the forming mandrel.  
The
modified Wolter/Giacconi design maintains a good psf over a wide field of
view.  Traditional designs such as XMM and JET-X have a psf that degrades
rapidly off axis.
\label{optical_perf}}
\end{figure}

The key technical requirement for our survey is to obtain a
sufficiently large grasp with sufficient angular resolution to
distinguish clusters from point sources throughout the field of view.
The requirements listed in Table 1 are met, for example, by a pair of
Wide Field X-ray Telescopes (WFXT) based on a modified Wolter/Giacconi
design that provides a high angular resolution of $\sim15$ arcsec  HEW
over a very large field of view of $\sim 1.4$ deg$^{2}$ which could be 
accomodated in a NASA Medium Explorer mission of 2 year duration.
To achieve the best design, we expand the mirror profiles in 2nd order polynomials along the optical axis (Burrows et al. 1992, Conconi \& Campana 2000).
The X-ray telescopes consist of 50 nested grazing incidence mirror shells each, with diameters ranging from 21 to 70 cm and a focal length of 3.5 m. The total mirror height is 28 cm.
The telescopes have effective area for various coatings is shown in
figure \ref{aeff}.
Possible technologies for high-performance 
high-throughput wide-field X-ray telescopes are ceramics, i.e. Silicon 
Carbide (SiC), or thin thermally-formed glass segments.
The feasibility of large wide-field SiC mirror shells has already been demonstrated by constructing two ceramic prototype shells of 60 cm diameter at OAB/Zeiss. The mirror shells have already been tested at the MPE/Panter and NASA/Marshall X-ray facilities showing imaging performance around 10-20 arcsec across a 60 arcmin field of view (Citterio et al. 1999; Ghigo et al. 1999)
as shown in figure \ref{optical_perf}. 
The measured performance is best at the lowest energies;  the degraded psf at
higher energies is understood, and is a consequence of increased roughness
on a replication mandrel which had been used many times.  Repolishing the
mandrel has resulted in significant improvement, although measurements have
only been made on-axis to date.
The slumped glass technology is being developed in the US and in Europe in the context of the Con-X/XEUS mission. 

\begin{figure}
\includegraphics[width=1.0\columnwidth,angle=0.0]{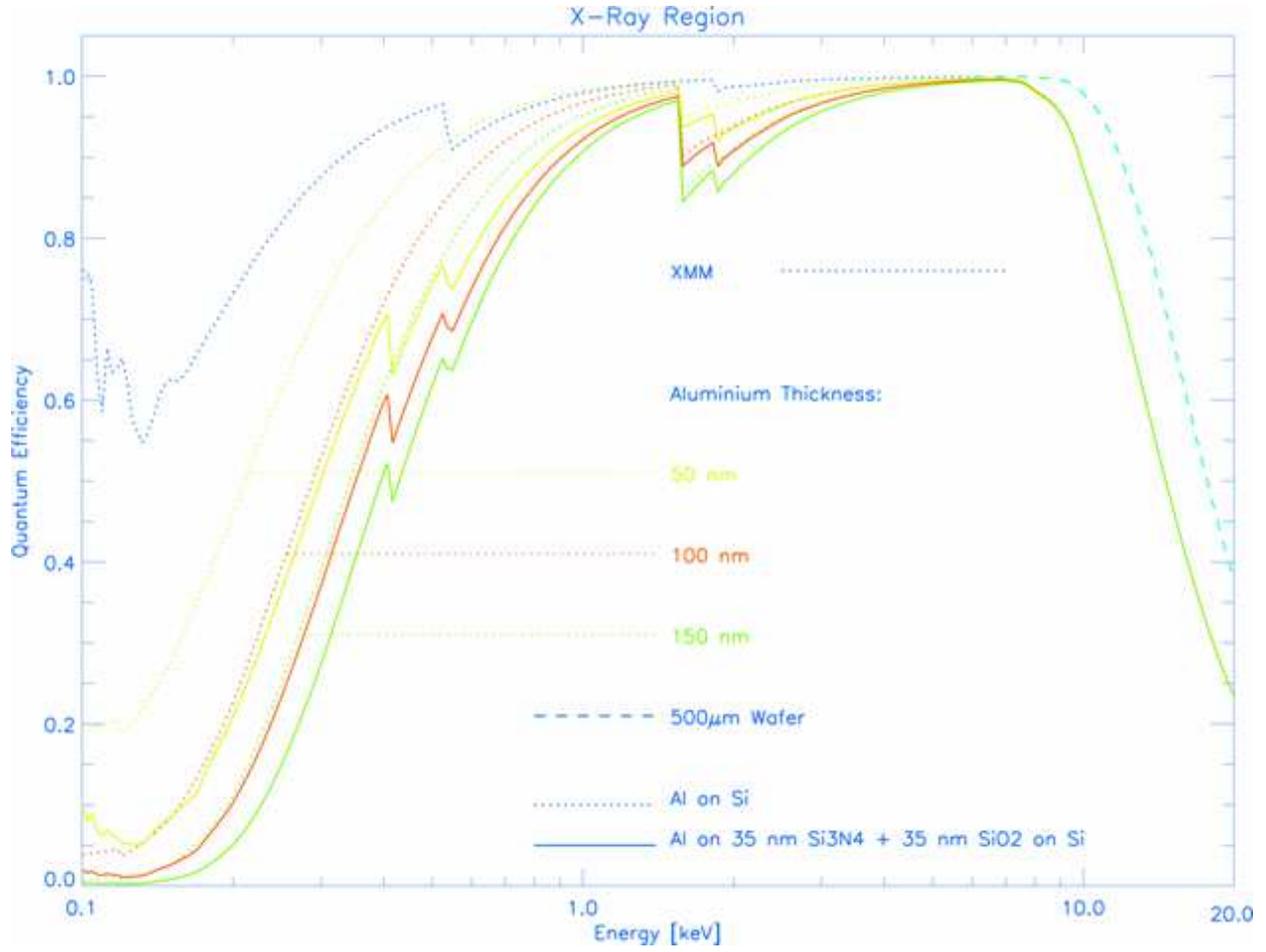}
\figcaption{The quantum efficiency of the pn-CCD chip based on
improvements to the XMM EPIC-pn CCD.
\label{ccd}}
\end{figure}

Current state-of-the-art, back-illuminated X-ray CCD technology is
fully capable of meeting the focal-plane instrument  requirements of
our survey.  Examples include pn-CCD detectors based on the XMM-Newton
and
DUO/ROSITA heritage, and back-illuminated Astro-E2 detectors with
Chandra/ACIS and HETE II heritage.  Individual
detectors would be mounted in an inverted pyramid to best conform to
the optimal focal surface, as has been done for
both Chandra/ACIS and XMM-Newton/EPIC.  The energy resolution and
readout speed capabilities of the current generation of
detectors, which can be operated at -40 to -60 deg C, are significantly
better than those flown on Chandra and XMM-Newton.
Directly-deposited optical blocking filters have been demonstrated for
these detectors, allowing very thin dead layers
and unprecedented quantum efficiencies across the 0.1-15 keV band.  For
example, the quantum efficiency of the
450-$\mu$m-thick, back-illuminated pn-CCD with a thin, directly
deposited aluminum/SiO blocking layer, is shown in
figure \ref{ccd}.

For our sensitivity calculations we assume a 5 keV cluster at z=0.5. 
The energy conversion factor (table \ref{survey_description}) assumes net (telescope
plus detector) effective area
 of 670 cm$^2$ and 55 cm$^2$ at 1.5 keV and 8 keV.
We propose a 
wide survey covering the whole extragalactic sky (20,000 deg$^2$) 
reaching a cluster flux limit of $2.3 \times 10^{-14}~{\rm erg}~{\rm cm}^{-2}~{\rm s}^{-1}$
in the 0.5--2 keV band.  We expect to detect 100,000 
clusters with the redshift distribution shown in figure \ref{zdist}.
With 50 or more photons per cluster and a point spread function better than
20 arcsec HEW we are able to separate clusters from point sources out to
redshifts of 1.5 finding more than 90\% of clusters.  Completeness rises
rapidly above this threshold while contamination (false clusters due to the
superposition of point sources) is estimated to be $\sim 1\%$.  

Other focal plane implementations, such as new developments in
cryogenic, large format spectroscopic arrays, could broaden the
scientific scope of this mission by providing the capability
to survey the large scale structure of the universe by tracing the
structure of the Warm/Hot Intergalactic Medium, and by providing
for many of the brighter clusters, independent X-ray derived redshifts.
Such capabilities, which are being independently developed for 
future observatories, are not required for the baseline mission
described here.


\clearpage

\clearpage
\begin{table}
\caption{Characteristics of the Survey. 
\label{survey_description}}
\vspace{\baselineskip}
\begin{tabular}{|l|rl|}
\hline
focal length & 350 & cm \\
Number of telescopes  &  2  & \\
On-axis effective area (@ 1 keV) & $\ge 670$ & cm$^2$ \\
Field of view                    & 1.4 & deg$^2$ \\
Vignetting factor                & 0.5 &         \\
Grasp (total for 2 telescopes)   & 940 & cm$^2$~deg$^2$ \\
PSF HEW\tablenotemark{a} (current)                & 20  & arcsec \\
PSF HEW (goal)                   & 15  & arcsec \\
ECF \tablenotemark{b} per 10$^{-11}$ (cgs) cluster & 7.9 & ct~sec$^{-1}$ \\
sensitivity                      & $2.3 \times 10^{-14}$ & ${\rm erg}~{\rm cm}^{-2}~{\rm s}^{-1}$ (0.5 -- 2 keV) \\
Cluster surface density          & 5.1 & deg$^{-2}$ \\
Exposure time per pointing                   & 2700 & sec \\
Number of pointings              & 14,000 &     \\
Solid angle                      & 20,000 & deg$^2$ \\
Number of clusters               & 100,000 &        \\
Mission duration                 & 2      & year   \\
Observing efficiency                       & 0.60     &        \\
\hline
\end{tabular}
\tablenotetext{a}{Point Spread Function Half Energy Width}
\tablenotetext{b}{Energy Conversion Factor ct s$^{-1}$(0.5 -- 10 keV)/erg (0.5 -- 2 keV) cm$^{-2}$ s$^{-1}$}
\end{table}

\clearpage

\end{document}